\documentclass[7pt,twoside,a4paper]{article}
\usepackage[affil-it]{authblk}
\usepackage{ amssymb }
\usepackage{amsmath}
\usepackage{color}
\usepackage{tikz}
\usetikzlibrary{matrix}
\usepackage[margin=2.9 cm]{geometry}
\usepackage{polski}
\usepackage[cp1250]{inputenc}
\usepackage[polish,english]{babel}

\input xy
\xyoption{all} \tolerance=500

\def\sT{\mathsf T}
\newdir{ (}{{}*!/-5pt/@^{(}}
\newdir{|>}{!/4,5pt/@{|}*:(1,-.2)@^{>}*:(1,+.2)@_{>}}

\def\mg{\mathfrak g}
\def\wta{\widetilde\alpha}
\def\wtb{\widetilde\beta}
\def\cob{\color{blue}}
\def\cor{\color{red}}

\title{The Tulczyjew triple in mechanics on a Lie group}
\author{Katarzyna Grabowska \thanks{Research supported by the  Polish National Science Centre grant under the contract number DEC-2012/06/A/ST1/00256}\\  Marcin Zając}
\affil{Faculty of Physics \\ University of Warsaw,\\
 Pasteura 5, 02-093 Warsaw, Poland }

\begin{document}

\maketitle

\begin{center}
\textbf{Abstract}
\end{center}
Tulczyjew triple for physical systems with configuration manifold equipped with Lie group structure is constructed and discussed. The case of systems invariant with respect to group acton is considered together with appropriate reduction of the Tulczyjew triple. The theory is applied to free rigid-body dynamics.

\begin{flushleft}
\textit{MSC 2010: 22E70, 53D05, 53Z05, 70E17, 70H33}
\end{flushleft}

\begin{flushleft}
\textit{Key words: Tulczyjew triple, Lie groups, classical mechanics, double vector bundles, rigid body}
\end{flushleft}

\section{Introduction}\label{sec:1}

Tulczyjew triple is a very useful commutative diagram built on maps that are essential in Lagrangian and Hamiltonian description of physical systems. In fact the name {\it Tulczyjew triple} refers not to one diagram but a collection of diagrams adapted to various physical situations. The very first triple introduced by Tulczyjew in his numerous works (e.g \cite{WMT3,WMT4,WMT6}) served for autonomous analytical mechanics. It was then adapted and generalised for time-dependent mechanics, mechanics on algebroids \cite{GGU,GG}, constrained mechanics on Dirac algebroids \cite{GG2}, for field theory \cite{G}, for higher order systems \cite{GV} etc. The concept of Tulczyjew triple points also to the certain philosophy of interpreting concepts of variational calculus within physical theories.

There exist many important physical systems with a configuration space which is a Lie group and with Lagrangian and Hamiltonian invariant under the group action. The most known is a free rotating rigid body but one can mention also a heavy top as well as infinite dimensional examples of ideal fluid and magnetohydrodynamics  \cite{HMR}. In such cases the dynamics may be reduced with respect to the group action leading to simplified equations with reduced number of degrees of freedom.

The aim of our paper is to construct the reduced version of Tulczyjew triple in finite dimensional case, when the configuration manifold is a Lie group. The Tulczyjew triple on a Lie group has not been widely considered in literature yet. Nevertheless, one may find some details in \cite{E,GR2,OE1,OE2}. In \cite{OE1,OE2} Essen and Gumral analyse a similar reduction of the Tulczyjew triple based on a semidirect product. In our paper we present a simpler way of such reduction without referring to semidirect products. On the other hand, there is an extensive literature about more general concept of mechanics on Lie algebroids initiated by Winstein \cite{W} and Liebermann \cite{LM2} and developed by many others, especially Martinez and his collaborators \cite{M1,M2,M3,M4}. As we have mentioned before, there exists also the carefully constructed Tulczyjew triple based on the structure of algebroid (not necesarily Lie algebroid) \cite{GGU,GG}. In this context our work fills a certain gap between classical Tulczyjew triple and its algebroidal generalization.

The paper is organized as follows. In section \ref{sec:2} we introduce the notation and basic geometrical background for our further work. In section \ref{sec:3} we recall three canonical isomorphisms of double vector bundles $\kappa_M$, $\alpha_M$, $\beta_M$ needed for classical Tulczyjew triple. In particular we present a way of constructing these maps that is later applied to a group-case situation. Section \ref{sec:4} contains a short description of the Tulczyjew triple and its role in Lagrangian and Hamiltonian mechanics. In section \ref{sec:5} we present a concise introduction to geometry of Lie groups especially tangent and cotangent bundles of a manifold that is a Lie group based on \cite{DK}. Next two sections contain main results of our paper. We construct the Tulczyjew triple over a manifold that is a Lie group. We use trivialized bundles $\sT\sT G$, $\sT\sT^*G$, $\sT^*\sT G$, $\sT^*\sT^*G$ and trivialised isomorphisms between them. We end section \ref{sec:6} with the diagram of trivialized Tulczyjew triple and its application to mechanics. In section \ref{sec:7} we consider the reduction of the Tulczyjew triple with respect to the group structure. We end this section with the diagram of reduced Tulczyjew triple and its application to mechanics. Finally, in section \ref{sec:8} we apply derived formalism to rigid body dynamics.

\subsection{Notation}\label{sec:2}

Let $M$ be a smooth manifold of dimension $m$. In an open subset $\mathcal{U}\subset M$ we introduce local coordinates $(q^i)^m_{i=1}$. The tangent and cotangent bundles will be denoted by $\tau_M: \mathsf TM\to M$ and $\pi_M: \mathsf T^\ast M\to M$ respectively. We have induced coordinates  $(q^i,\dot q^j)$ in $\tau_M^{-1}(\mathcal{U})$. Thus, we can write $\tau_M$ as
$$\tau_M : \sT M\ni (q^i,\dot{q}^j)\longmapsto  (q^i)\in M.$$
The tangent bundle plays the role of a set of positions and velocities of a mechanical system. It is also called a {\it space of infinitesimal configurations}. Let $\gamma:\mathbb{R}\supset I\ni t\longmapsto \gamma(t)\in M$ be a smooth curve. The tangent vector to this curve at point $t_0\in I$ will be denoted by $\dot\gamma(t_0)$ or equivalently $\frac{\mathrm{d}}{\mathrm{dt}}_{t=0}\gamma$.

Local induced coordinates in $\pi_M^{-1}(\mathcal{U})\subset \sT^\ast M$ will be denoted by $(q^i,p_j)$. The projection $\pi_M$ may be locally written as
$$\pi_M : {\mathsf T}^*M\ni (q^i,p^j)\to (q^i) \in M.$$
The cotangent bundle represents a set of positions and momenta of a physical system and is also called {\it phase space} of a system.

We introduce induced coordinates in iterated tangent and cotangent bundles
\begin{eqnarray*}
(q^i,\dot{q}^j,\delta q^k, \delta \dot{q}^l) &\text{in}& \sT\sT M, \\
(q^i,p_j,\dot{q}^k,\dot{p}_l) &\text{in}&   \sT\sT^*M, \\
(q^i,\dot{q}^j,p_k, \phi^l) &\text{in}& \sT^*\sT M, \\
(q^i,p_j,\pi_k, x^l) &\text{in}&  \sT^*\sT^*M.
\end{eqnarray*}
The above four bundles play a crucial role in our paper. They have rich geometric structure. Each of them is a {\it double vector bundle} i.e. it has two compatible structures of a vector bundle (see \cite{GR},\cite{P},\cite{KU}).
One can show that there exist a canonical isomorphism of vector bundles $\sT^*\sT^*M$ and $\sT^*\sT M$ denoted by $\gamma_M$ \cite{KU,GU2}. In fact such an isomorphism of $\sT^\ast E^\ast$ and $\sT^\ast E$ exists for any vector bundle $E$. In coordinates, for $E=\sT M$ it reads
\begin{equation}\label{gamma}
\gamma_M:\sT^*\sT^*M\to\sT^*\sT M, \quad (q^i,p_j,\pi_k, x^l)\longmapsto (q^i, x^l,-\pi_k, p_j).
\end{equation}
The following projection $\zeta_{\sT M}$ is a part of the double vector bundle structure of $\sT^\ast\sT M$ \cite{KU,GU2}.
\begin{equation}\label{zeta}
\zeta_{\sT M}: \sT^\ast\sT M\to\sT^\ast M, \quad (q^i,\dot{q}^j,p_k, \phi^l)\longmapsto (q^i, \phi^l).
\end{equation}
The cotangent bundle has a natural structure of a symplectic manifold denoted by $(\sT^*M,\omega_M)$ \cite{LM1}. The symplectic form is the differential of the canonical {\it Louville form} defined as
$$\theta_M:\sT\sT^*M \ni v\to \langle \tau_{\sT^*M}(v),\sT\pi_M(v)\rangle \in\mathbb R,$$
where $\tau_{\sT^*M}$ and $\sT\pi_M$ are two structures of a vector bundle in $\sT\sT^*M$. In coordinates we can write
\begin{equation}\label{forma}
\theta_M=p_i\mathsf dq^i, \qquad \omega_M:=\mathsf d\theta_M= \mathsf dp_i\wedge\mathsf dq^i.
\end{equation}

\subsection{Canonical isomorphisms of iterated tangent and cotangent bundles}\label{sec:3}
In this section we will discuss some particular double vector bundles i.e. tangent and cotangent bundles of $\mathsf TM$ and $\sT^*M$. In particular we will consider three isomorphisms of these double vector bundles that form the classical version of the Tulczyjew triple.

Double vector bundle $\sT\sT M$ may be represented by the following diagram
\begin{equation}\label{ttm}
\xymatrix {
&{\mathsf T}{\mathsf T}M \ar[dl]_{\tau_{{\mathsf T}M}}\ar[dr]^{{\mathsf T}\tau_M}& \\
{\mathsf T}M \ar[dr]^{\tau_M}& &{\mathsf T}M \ar[dl]_{\tau_M} \\
&M&
}
\end{equation}
The two vector bundle projections in coordinates read
\begin{align*}
\tau_{{\mathsf T}M} &: \sT\sT M\ni (q^i,\dot{q}^j,\delta q^k, \delta \dot{q}^l)\to (q^i,\dot{q}^j)\in {\mathsf T}M,\\
{\mathsf T}\tau_{M} &: {\mathsf T}{\mathsf T}M\ni (q^i,\dot{q}^j,\delta q^k, \delta \dot{q}^l)\to (q^i,\delta q^k)\in {\mathsf T}M.
\end{align*}
It is well known that there exists the canonical isomorphism
\begin{equation}\label{kappa}
\kappa_M: \sT\sT M\longrightarrow \sT\sT M, \quad (q^i,\dot{q}^j,\delta q^k,\delta \dot{q}^l)\longmapsto (q^i,\delta q^k,\dot{q}^j,\delta \dot{q}^l).
\end{equation}
Let us recall the definition of $\kappa_M$, because we shall need it later for $M$ being a Lie group. Each element of $\mathsf T\mathsf TM$ is an equivalence class of homotopies $\mathbb R^2\ni (s,t)\to \chi (s,t)\in M,$ i.e.
$$v=\frac{\mathsf d}{\mathsf ds}_{|{s=0}}\frac{\mathsf d}{\mathsf dt}_{|t=0}\chi. $$
The map $\kappa_M$ is defined on representatives. If $\chi(s,t)$ is a homotopy from class $v$ then $\kappa_M(v)$ is the equivalence class of a homotopy $\bar\chi(s,t):=\chi(t,s)$, i.e. $\chi$ with flipped arguments. We have then
\begin{equation}\label{w:1}
\kappa_M : \frac{\mathsf d}{\mathsf ds}_{|{s=0}}\frac{\mathsf d}{\mathsf dt}_{|t=0}\chi\longmapsto\frac{\mathsf d}{\mathsf ds}_{|{s=0}}\frac{\mathsf d}{\mathsf dt}_{|t=0}\bar\chi.
\end{equation}
It is clear from the definition that $\kappa_M$ interchanges two vector bundle structures in $\sT\sT M$. It is an isomorphism of double vector bundles with the following diagram
\begin{equation}\label{diag:2}
\xymatrix{
\sT\sT M \ar[rr]^{\kappa_M} \ar[d]_{\tau_{\sT M}} & &\sT\sT M \ar[d]^{\sT\tau_M} \\
\sT M \ar@{=}[rr] & & \sT M
}
\end{equation}

The {\it Tulczyjew isomorphism} $\alpha_M$ which constitutes the Lagrangian side of the Tulczyjew triple is dual to the above vector bundle isomorphism $\kappa_M$ (see e.g. \cite{WMT1}). It is clear that the bundle $\pi_{{\mathsf T}M}:{\mathsf T}^*{\mathsf T}M\to {\mathsf T}M$ is dual to the bundle $\tau_{{\mathsf T}M}$. The bundle ${\mathsf T}\pi_{M} : {\mathsf T}{\mathsf T}^*M\to {\mathsf T}M$ is dual to ${\mathsf T}\tau_{M}$ with respect to the pairing
\begin{equation}\label{e:6}
\langle\!\langle\rho,v\rangle\!\rangle=\frac{\mathsf d}{\mathsf dt}_{|t=0}\langle\xi(t),\gamma(t)\rangle,
\end{equation}
where $v\in\sT\sT M$ and $\rho\in\sT\sT^*M$ are tangent vectors such that ${\sT\tau_M}(v)={\sT\pi_M}(\rho)$, and $\gamma,\xi$ are curves
\begin{equation}
\gamma:\mathbb R\ni t\to\sT M, \quad \xi:\mathbb R\ni t\to \sT^*M,
\end{equation}
representing $v$ and $\rho$ respectively, chosen in such a way that they have the same projection on $M$. The evaluation $\mathbb R\ni t\to \langle\xi(t),\gamma(t)\rangle$ is therefore well-defined. Let us notice that $\langle\!\langle\cdot,\cdot\rangle\!\rangle$ is nondegenerate. In coordinates for $v=(q^i,\delta q^k,\dot{q}^j,\delta\dot{q}^l)$ and $\rho=(q^i,p_j,\dot{q}^k,\dot{p}_l)$ (\ref{e:6}) reads
$$\langle\!\langle\rho,v\rangle\!\rangle=\delta q^j\dot p_l+\delta\dot q^kp_j. $$
Using the pairing above and formula (\ref{kappa}) one can easily find $\alpha_M$ in coordinates
$$\alpha_M:\sT\sT^* M\to\sT^*\sT M,\quad (q^i,p_j,\dot{q}^k,\dot{p}_l)\longmapsto (q^i,\dot{q}^k,\dot{p}_l,p_j).$$

The map $\alpha_M$ is an isomorphism of double vector bundles. It is also a symplectomorphism with respect to forms $\omega_{\sT M}$ and $\mathrm{d}_{\sT}\omega_M$ \cite{GU,PU,WMT1,WMT7}. The form $\mathrm{d}_\sT\omega_M$ is a canonical symplectic form on $\sT\sT^\ast M$ -- a tangent lift of the canonical symplectic form on $\sT^*M$. In natural coordinates we have
$$\mathrm{d}_\sT\omega_M=\mathrm{d}\dot p_i\wedge\mathrm{d}q^i+\mathrm{d}p_j\wedge\mathrm{d}\dot q^j.$$

The third important isomorphism in the context of our paper is the map $\beta_M$ derived from the symplectic form on $\sT^*M$
$$\beta_M:  \sT\sT^\ast M\rightarrow\sT^\ast\sT^\ast M, v\longmapsto \omega_M(\cdot,v). $$
It is an isomorphism of double vector bundles $\mathsf T\mathsf T^*M$ and $\mathsf T^*\mathsf T^*M$ and antisymplectomorphism with respect to $\omega_{\sT^\ast M}$ and $\mathrm{d}_{\sT}\omega_M$. In coordinates it reads
\begin{equation}\label{beta:2}
\beta_M:\sT\sT^* M\to\sT^*\sT^* M,\quad (q^i,p_j,\dot{q}^k,\dot{p}_l)\longmapsto(q^i,p_j,-\dot p_l,\dot q^k).
\end{equation}
The above map can be used to produce vector fields on the cotangent bundle from functions on it. For a smooth function $f:\mathsf T^*M\to\mathbb R$ we define the vector field $X_f$ composing $\mathrm{d} f$ with $\beta_M$
$$X_f:\sT^*M\to\sT\sT^*M, \quad p\longmapsto\beta_M^{-1}\circ\mathrm df(p), $$
$$X_f=\frac{\partial f}{\partial p_k}\partial_{q^k} -\frac{\partial f}{\partial q^l}\partial_{p_l}. $$
The vector field $X_f$ is called {\it the Hamiltonian vector field} of the function $f$.

Let us notice, that once we have one of the maps $\alpha_M$, $\beta_M$, $\kappa_M$ we can define the other two using canonical structures present on any vector bundle $E$. Maps $\kappa_M$ and $\alpha_M$ are connected by duality while $\beta_M$ is related to $\alpha_M$ by $\gamma_M$ of formula (\ref{gamma}). It means that each of these three mappings contains the same information.

\subsection{The Tulczyjew Triple}\label{sec:4}
We will now present a point of view on describing a mechanical system which is alternative to the one present in most textbooks \cite{K}. Its essence lies in the so-called {\it Tulczyjew triple}. Tulczyjew triple enables us to describe systems in both Lagrangian and Hamiltonian approach and shows relation between the two. It is important to notice, that the triple is based only on {\it canonical} structures of proper bundles. We will not derive here the whole formalism whose origin lies in reinterpretation of variational description of statical systems. One can find the thorough analysis with all the details in numeorus Tulczyjew papers e.g. \cite{WMT1,WMT2,WMT3,WMT4,WMT5,WMT6}.

The Tulczyjew triple is a geometrical structure presented in the followig diagram

\begin{equation}\label{t:9}
\xymatrix{
& & & \color{red}{D} \ar@{^{(}->}[d]&& \\
& {\mathsf T}^*{\mathsf T}^*M \ar[rd]^{\pi_{\mathsf T^*M}}& & {\mathsf T}{\mathsf T}^*M\ar[ll]_{\beta_M}\ar[rr]^{\alpha_M}\ar[ld]_{\tau_{\mathsf T^*M}}\ar[rd]^{\sT\pi_M} & &{\mathsf T}^*{\mathsf T}M \ar[ld]_{\pi_{\mathsf TM}} \\
 & &{\mathsf T}^*M \ar[rd] \ar@/_1pc/[ur]_{{\color{red}X_H}}  \ar@/^1pc/[ul]^{{\color{red}{\mathsf d}H}}& & {\mathsf T}M \ar[ld] \ar@/_1pc/[ur]_{\color{red}{{\mathsf d}L}}&&  \\
&&&M&&&
}
\end{equation}
The right-hand side of the diagram is related to Lagrangian formalism while the left-hand side to Hamiltonian one. Both formalisms are based on the same scheme and the only difference is in generating objects on both sides. We will discuss now more precisely Lagrangian and Hamiltonian description in the language of Tulczyjew triple. For simplicity, we will consider only autonomous systems i.e. with no external forces.

Let $M$ be the configuration manifold of the system and $L:\sT M\longrightarrow \mathbb{R}$ its Lagrangian. {\it The dynamics} of the system is a subset
\begin{equation}\label{t:2}
\mathcal D:=\alpha_M^{-1}\circ\mathrm dL(\sT M)
\end{equation}
of $\sT\sT^*M$. Since $\mathsf dL(\sT M)$ is a Lagrangian submanifold in $\sT^*\sT M$, and $\alpha_M$ is a symplectomorphism, the dynamics is a Lagrangian submanifold in $\sT\sT^*M$. From physical point of view it is a (possibly implicit) first order differential equation for a trajectory in the phase space. A curve \mbox{$\eta: \mathbb{R}\supset I\rightarrow\sT^\ast M$} is a phase trajectory if $\dot\eta(t)\in \mathcal{D}$ for $t\in I$.

Now let us assume that Hamiltonian $H:\sT^\ast M\rightarrow\mathbb{R}$ of the system exists. The dynamics of the system is then the image of the Hamiltonian vector field $X_H$ i.e.
\begin{equation}\label{t:21}
\mathcal{D}=X_H(\sT^\ast M)=\beta_M^{-1}(\mathrm{d}H(\sT^\ast M)).
\end{equation}
Phase space trajectories are integral curves of the field $X_H$.

The dynamics of the system may be projected on $\sT M\times \sT^*M$. The projection $\Lambda=\sT\pi_{M}\times \tau_{\sT^\ast M}(\mathcal{D})$ is a subset of Cartesian product of $\sT M$ and $\sT^*M$ therefore it can be understood as a relation between these two manifolds. If the dynamics comes from Lagrangian it is a graph of the Legendre map
$\lambda=\zeta_M\circ\mathrm{d}L$. If the dynamics comes from Hamiltonian it is a graph of a map `in opposite direction'. In general, it can be a relation that is not a map at all. In such a case in place of Lagrangian and Hamiltonian we have to consider more general objects generating Lagrangian submanifolds than functions on  manifolds \cite{LM1,WMT7}.

\subsection{The tangent and cotangent bundle of a Lie group}\label{sec:5}

In this section we shall briefly recall some basic information about geometry of Lie groups. One can find the detailed description of this subject in \cite{DK}. Let $(G, \cdot)$ be a Lie group. For each $g\in G$ we introduce left and right translations
\begin{equation}\label{e:2}
\begin{aligned}
&l_g:G\to G :l_g(x)=gx, \\
&r_g:G\to G :r_g(x)=xg.
\end{aligned}
\end{equation}
Both maps are diffeomorphisms. In our constructions, according to the tradition, we shall rather use left translation. The left and right translation maps define a family of group automorphisms
$$\mathrm{Ad}_{g}: G\to G,\quad x\longmapsto l_{g}\circ r^{-1}_{g}(x)=gxg^{-1}. $$
The tangent map $\sT l_{g^{-1}}$ restricted to $\sT_g G$ is a linear isomorphism between $\sT_g G$ and $\sT_eG=\mathfrak{g}$. We can therefore trivialise the bundle $\sT G$ as follows
\begin{equation}\label{e:3}
\imath: \sT G\longrightarrow G\times \mathfrak{g}, \quad v_g\longmapsto (g, \sT l_{g^{-1}}(v_g)).
\end{equation}
The trivialized version of the tangent bundle of a Lie group is a bundle $pr_1: G\times \mg\rightarrow G$.
Each element $X\in\mathfrak{g}$ defines a vector field $X^{l}$ on a group $G$ given by $X^l(g):=\sT l_g(X)$. In the following we shall use the notation $gX:=\sT l_g(X)$. The field $X^l$ is by definition left invariant i.e. $X^l(l_g(x))=\sT l_g(X(x))$. The flow of a left invariant vector field will be denoted by $\phi^X_t$. The exponential map exp$:\mathfrak g\to G$ is given by the flow at $e$
\begin{equation}\label{e:5}
\exp(tX)=\phi^X_t(e).
\end{equation}
For matrix groups it is equivalent to a well known matrix exponential defined by a power series.

The tangent space $\mg$ is equipped with the the bracket $[X,Y]_{\mathfrak{g}}:=~[X^l, Y^l](e)$ therefore it has a natural structure of a Lie algebra. In the following we will skip the index $\mathfrak{g}$ in $[\cdot,\cdot]_{\mathfrak g}$ if it does not lead to any confusion. The Lie bracket defines a map $\mathrm{ad}_{X}:\mathfrak{g}\to\mathfrak{g}$, ${Y}\longmapsto
[X,Y]$. There is a well known relation between $\mathrm{Ad}$ and $\mathrm{ad}$ maps
$$\mathrm{ad}_X=\frac{\mathsf d}{\mathsf dt}_{|{t=0}}\mathrm{Ad_{exp(tX)}}.$$
Now let $\sT^*G$ be the cotangent bundle of $G$. Trivialisation $\sT G\simeq G\times\mathfrak g$ by $\imath$ leads to the trivialisation of $\sT^*G$, namely $\sT^*G\simeq G\times\mathfrak g^*$ where $\mathfrak g^*$ is dual to $\mathfrak g$. More precisely we use $\xi:=(\imath^*)^{-1}$
$$\xi :\sT^*G\to G\times \mathfrak g^*,\quad b_g\longmapsto (g,(\sT l_g)^\ast(b_g)),$$
where $b_g$ is a tangent covector at point $g$. We will denote elements of $\mg^\ast$ by capital letters from the beginning of the alphabet while elements of $\mg$ will be denoted by capital letters from the end of the alphabet.

Once we have trivializations $\imath$ and $\xi$ we can trivialize any iterated tangent and cotangent bundle. We will use the fact that if $V$ is a vector space then $\sT V\simeq V\times V$ and $\sT^* V\simeq V\times V^*$.

\section{Tulczyjew Triple on a Lie group}\label{sec:6}

One of the advantages of the Tulczyjew triple is the posibility of reducing it with respect to symmetries of the system. In this section we will consider a symmetry associated to the group action on itself. Our task is to find a trivialized form of Tulczyjew triple in the case $M=G$ and then reduce it with respect to the symmetry. In particular, we have to find trivialized isomorphisms $\kappa_G$, $\alpha_G$ \mbox{and $\beta_G$.}

\subsection{Canonical involution of iterated tangent bundle on a Lie group}

Using trivialization $\imath$ we get ${\mathsf T}{\mathsf T}G\simeq {\mathsf T}(G\times \mathfrak{g})\simeq G\times \mathfrak{g}\times \mathfrak{g}\times \mathfrak{g}$. The diagram representing a double vector bundle $G\times \mathfrak{g}\times \mathfrak{g}\times \mathfrak{g}$ reads
$$\xymatrix{
&G\times\mathfrak{\color{red}{g}}\times\mathfrak{\color{blue} g}\times\mathfrak g \ar[dl]_{pr_{12}} \ar[dr]^{pr_{13}}& \\
G\times\mathfrak{\color{red}g}\ar[dr]^{pr_1} & &G\times\mathfrak{\color{blue} g} \ar[dl]_{pr_1} \\
&G&
}$$
where $pr_{12}$ refers to $\tau_{\sT G}$ and $pr_{13}$ refers to $\sT\tau_G$ (see diagram (\ref{ttm})).

Let $v=(g,X,Y,Z)\in G\times \mg\times\mg\times\mg$ be a trivialized element of $\sT\sT G$ at point $(g,X)$ (i.e. in \mbox{$gX\in\sT G$)}.
In the following we shall not distinguish between elements of $\sT\sT G$ (and other iterated tangent and cotangent bundles of the group) and their trivialized versions. It means, that by \mbox{$v=(g,X,Y,Z)$} we will understand an element of $\sT\sT G$, such that after trivialization it takes the above form. The homotopy representing vector $v$ can be written as
\begin{equation*}
\chi:\mathbb{R}^2\to G,\quad (s,t)\longmapsto g\exp(sY)\exp(tX+stZ).
\end{equation*}
Indeed, one can easily check that
\begin{equation*}
\frac{\mathsf d}{\mathsf ds}_{|{s=0}}\frac{\mathsf d}{\mathsf dt}_{|t=0}\chi=(g,X,Y,Z)\in G\times\mathfrak g\times\mathfrak g\times\mathfrak g.
\end{equation*}
Then $\bar\chi$ reads
\begin{equation*}
\bar\chi(s,t)=g\exp(tY)\exp(sX+stZ)\in G.
\end{equation*}
Differentiating $\bar\chi$ with respect to $t$ at $t=0$ and  $s$ at $s=0$ respectively we get
\begin{equation*}
\frac{\mathsf d}{\mathsf ds}_{|{s=0}}\frac{\mathsf d}{\mathsf dt}_{|t=0}\bar\chi=(g,Y,X,-\mathrm{ad}_{X}Y+Z)=(g,Y,X,Z-[X,Y])\in G\times\mathfrak g\times\mathfrak g\times\mathfrak g.
\end{equation*}
Finally, the trivialized $\kappa_G$ has the following form
\begin{equation}\label{l:1}
\xymatrix{
\textcolor{red}{G\times \mathfrak{g}}\times \mathfrak{g}\times \mathfrak{g} \ar[rr]^{\widetilde\kappa_G} \ar[d]_{pr_{12}} & &\textcolor{blue}{G}\times \mathfrak{g}\times \textcolor{blue}{\mathfrak{g}}\times \mathfrak{g} \ar[d]^{pr_{13}} \\
\textcolor{red}{G\times \mathfrak{g}} \ar@{=}[rr] & &\textcolor{blue}{G\times \mathfrak{g}}
}
\end{equation}
\begin{equation}\label{kappa2}
\widetilde\kappa_G:G\times\mathfrak{g}\times\mathfrak{g}\times\mathfrak{g}\to G\times\mathfrak{g}\times\mathfrak{g}\times\mathfrak{g},\quad (g,X,Y,Z)\longmapsto (g,Y,X,Z-[X,Y]).
\end{equation}

\subsection{Bracket of vector fields on a Lie group}

In this section we shall derive a formula for trivialized bracket of vector fields on $G$. Composing a vector field $\xi:G\to\sT G$ with $\imath$ we get a trivialized vector field
$$\imath\circ \xi:G\ni g\to (g,X(g))\in G\times\mathfrak g.$$
where $X(g)=\sT l^{-1}_g(\xi(g))$. One can now ask the following question: if $\xi,\eta$ are vector fields on $G$, then what is a trivialized version of a vector field $[\xi,\eta]$, i.e. what is the value of $\sT l^{-1}_g([\xi,\eta](g))$? If $\xi$ and $\eta$ are left invariant then the answer to this question is trivial and comes from the definition of Lie structure in $\mathfrak g$. Let us then consider the case when $\xi$ and $\eta$ are in general not left invariant vector fields.

We will start with a useful formula that one can easily prove. Let $\xi,\eta$ be vector fields on $G$. Then we have
\begin{equation}\label{l:10}
\kappa_G(\sT \eta(\xi))-\sT \xi(\eta)=[\xi,\eta]^{\mathsf v}_\xi,
\end{equation}
where $[\xi,\eta]^{\mathsf v}_\xi$ is a vertical lift of $[\xi,\eta]$ to $\xi$. More precisely, $[\xi,\eta]^{\mathsf v}_\xi$ is an element tangent to the curve
$$\mathbb R\ni t\to \xi(g)+t[\xi,\eta](g)\in\sT G.$$
The formula is well known in the literature, but it is unclear who was the first one to use it. It holds for any smooth manifold not just a Lie group. Let the trivialized versions of fields $\xi,\eta$ be as follows
$$\imath\circ\xi=(g,X(g))\in G\times\mathfrak g, \qquad \imath\circ\eta=(g,Y(g))\in G\times\mathfrak g,$$
It turns out that the trivialized version of (\ref{l:10}) is
\begin{equation}
pr_2\circ\imath([\xi,\eta](g))=pr_2\circ\sT Y(\sT l_gX(g))-pr_2\circ\sT X(\sT l_g(Y(g)))+[X(g),Y(g)]_{\mathfrak g}.
\end{equation}
If we put $DX:=pr_2\circ\sT X\circ\sT l_g$, $DY:=pr_2\circ\sT Y\circ\sT l_g$ and denote $pr_2\circ\imath([\xi,\eta](g))$ by $[X,Y](g)$ we can write
\begin{equation}\label{l:4}
[X,Y](g)=DY(X(g))-DX(Y(g))+[X(g),Y(g)]_{\mathfrak g}.
\end{equation}

\subsection{Tulczyjew isomorphism on a Lie group}
The trivialized version of $\alpha_G$ can be obtained from trivialized $\kappa_G$ by duality. For that we need trivialized bundles $\sT^\ast\sT G$ and $\sT\sT^\ast G$. The diagram representing the trivialised double vector bundle \mbox{$\sT\sT^*G\simeq G\times\mathfrak{g}^*\times\mathfrak{g}\times\mathfrak{g}^*$} reads
\begin{equation}\label{diag:3}
\vcenter{
$$\xymatrix@C-10pt{
&G\times\mathfrak{\color{red}{g^*}}\times\mathfrak{\color{blue} g}\times\mathfrak g^* \ar[dl]_{pr_{12}} \ar[dr]^{pr_{13}}& \\
G\times\mathfrak{\color{red}g^*}\ar[dr]^{pr_{1}} & &G\times\mathfrak{\color{blue} g} \ar[dl]_{pr_{1}} \\
&G& }
$$
}
\end{equation}
where $pr_{12}$ and $pr_{13}$ correspond to $\tau_{\sT^\ast G}$ and $\sT\pi_G$ respectively. For \mbox{$\sT^*\sT G\simeq G\times\mathfrak{g}\times\mathfrak{g}^*\times\mathfrak{g}^*$} we get
$$\xymatrix{
&G\times\mathfrak{\color{red}{g}}\times\mathfrak{\color{blue} g^*}\times\mathfrak g^* \ar[dl]_{pr_{12}} \ar[dr]^{pr_{14}}& \\
G\times\mathfrak{\color{red}g}\ar[dr]^{pr_{1}} & &G\times\mathfrak{\color{blue} g^*} \ar[dl]_{pr_{1}} \\
&G& }
$$
with $pr_{12}$ and $pr_{14}$ corresponding to $\tau_{\sT G}$ and $\zeta_{\sT G}$ respectively.

Since $\alpha_G$ is dual to $\kappa_G$ the same holds for trivialized versions $\widetilde\alpha_G$ and $\widetilde\kappa_G$. The diagram dual to (\ref{l:1}) reads
\begin{equation}\label{l:5}
\xymatrix{
\textcolor{red}{G\times \mathfrak{g}}\times \mathfrak{g^*}\times \mathfrak{g^*}  \ar[d]_{pr_{12}} & &\textcolor{blue}{G}\times \mathfrak{g^*}\times \textcolor{blue}{\mathfrak{g}}\times \mathfrak{g^*} \ar[d]^{pr_{13}} \ar[ll]_{\widetilde\alpha_G}\\
\textcolor{red}{G\times \mathfrak{g}} \ar@{=}[rr] & &\textcolor{blue}{G\times \mathfrak{g}}
}
\end{equation}
The pairing between trivialised versions of $\sT\sT G$ and $\sT^\ast\sT G$ treated as vector bundles over $\sT G$
is obvious. For $(g,X,Y,Z)\in\sT\sT G\simeq {G\times \mathfrak{g}}\times \mathfrak{g}\times \mathfrak{g}$ and $(g,X,A,B)\in \sT^\ast\sT G\simeq {G\times \mathfrak{g}}\times \mathfrak{g^*}\times \mathfrak{g^*}$ we get
$$\langle(g,X,A,B),\;\; (g,X,Y,Z)\rangle= \langle A,Y\rangle+\langle B,Z\rangle.$$
For $(g,A,Y,B)\in\sT\sT^\ast G\simeq G\times \mathfrak{g^*}\times\mathfrak{g}\times \mathfrak{g^*}$ and $(g,X,Y,Z)\in \sT\sT G\simeq{G}\times \mathfrak{g}\times\mathfrak{g}\times \mathfrak{g}$ we obtain the formula
$$\langle\!\langle(g,A,Y,B),\;\; (g,X,Y,Z)\rangle\!\rangle= \langle A,Z\rangle+\langle B,X\rangle,$$
which is a trivialized version of the pairing (\ref{e:6}). Finally, for $\widetilde\alpha_G$ being the trivialized $\alpha_G$ we have
\begin{equation}\label{l:8}
\widetilde\alpha_G: G\times \mathfrak{g^*}\times\mathfrak{g}\times \mathfrak{g^*}\to G\times \mathfrak{g}\times \mathfrak{g^*}\times \mathfrak{g^*}, \quad
(g,A,X,B)\longmapsto (g,X,B-\mathrm{ad}^*_X(A),A).
\end{equation}
In the end, we shall write the formula for $\widetilde\alpha_G^{-1}$ that will be more useful than $\widetilde\alpha_G$ itself
\begin{equation}\label{l:9}
\widetilde\alpha_G^{-1}(g,X,C,D)=(g,D,X,C+\mathrm{ad}^*_X(D)).
\end{equation}

\subsection{Symplectic form on the cotangent bundle of a Lie group}
As we stated, the $\beta_G$ is an isomorphism of bundles $\sT\sT^*G$ and $\sT^*\sT^*G$ associated to the symplectic form on $\sT^\ast G$. The trivialization of $\sT^*\sT^*G\simeq\sT^*(G\times\mg^*)$ gives $G\times\mathfrak{g^*}\times\mathfrak{g^*}\times\mathfrak{g}$ with the diagram
$$\xymatrix{
&G\times\mathfrak{\color{red}{g^*}}\times\mathfrak g^*\times{\color{blue}\mathfrak g} \ar[dl]_{pr_{12}} \ar[dr]^{pr_{14}}& \\
G\times\mathfrak{\color{red}g^*}\ar[dr]^{pr_{1}} & &G\times\mathfrak{\color{blue} g} \ar[dl]_{pr_{1}} \\
&G& }
$$
Projection $pr_{12}$ is a trivialised version of $\pi_{\sT^*G}$ while $pr_{14}$ is a trivialised version of  $\zeta_{\sT^*G}$.

The trivialized Louville form on $G\times\mathfrak g^*$ will be denoted by $\widetilde\theta_G$. Using the diagram (\ref{diag:3}) we get
$$\widetilde\theta_G:G\times\mathfrak g^*\times\mathfrak g\times\mathfrak g^*\ni w\to \langle pr_{12}(w),pr_{13}(w)\rangle \in\mathbb R,$$
which means that taking $w=(g,A,X,B)$ we obtain
$$\widetilde\theta_G(g,A,X,B)=\langle (g,A),(g,X)\rangle=\langle A,X\rangle. $$
Let $\Phi,\Psi$ be vector fields on $\sT^*G$, such that in trivialization they have the following form
$$\phi(g,A)=(g,A,X(g,A),B(g,A)), \qquad \psi(g,A)=(g,A,Y(g,A),C(g,A)). $$
Let us notice, that $X,B$ and $Y,C$ depend on $g$ and $A$ and they do not have to be left invariant. To find the trivialised version of $\omega_G$ we shall use the Cartan formula for differential
\begin{equation}\label{C1}
\omega_G(\Phi,\Psi)=\mathsf d\theta_G(\Phi,\Psi)=\Phi\theta_G(\Psi)-\Psi\theta_G(\Phi)-\theta_G([\Phi,\Psi]),
\end{equation}
and the following notation for derivations. For $Y:G\times \mathfrak{g}^\ast\rightarrow \mathfrak{g}$ and $X\in
\mathfrak{g}$, $D_X Y$ is a derivation of $Y$ with respect to the first argument and in the direction of $X$, more precisely
$$(D_XY)(g,A)=pr_2\circ\sT Y(g,A,X,0).$$
For $B\in \mathfrak{g}^\ast$ we write $D_BY$ for the derivation with respect to the second argument and in the direction of $B$, i.e.
$$(D_BY)(g,A)=pr_2\circ\sT Y(g,A,0,B).$$
Applying trivialization to first two elements from (\ref{C1}) means replacing $\Phi$ and $\Psi$ be $\phi$ and $\psi$. We get then
$$\phi\widetilde\theta_G(\psi)=\\
\langle A,(D_XY)(g,A)\rangle+\langle B,Y\rangle+\langle A,(D_BY)(g,A)\rangle,$$
similarly
$$\psi\widetilde\theta_G(\phi)=
\langle A,(D_YX)(g,A)\rangle+\langle C,X\rangle+\langle A,(D_CX)(g,A)\rangle. $$
The third component in formula (\ref{C1}), after trivialisation, is $\widetilde\theta_G([\phi,\psi])$. Let us rewrite $\phi$ as \mbox{$X(g,A)+B(g,A)$}, where $X$ and $B$ are components of the field $\phi$ in two different directions. We will do the same for $\psi$. The bracket of fields $\phi$ and $\psi$ may be written as
$$[\phi,\psi]=[X+B,Y+C]=[X,Y]+[X,C]+[B,Y]+[B,C].$$
To express the first component we use the formula (\ref{l:4})
$$[X,Y]=D_XY-D_YX+[X,Y]_{\mg}  $$
The next terms are
\begin{align*}
[X,C]&=D_XC-D_CX, \\
[B,Y]&=D_BY-D_YB, \\
[B,C]&=D_BC-D_CB, \\
\end{align*}
where we apply the same notation for differentiating with respect to the first and second arguments of $C$ and $B$.
In all the above terms only $D_BY$ and $D_CX$ are components along $\mg$, so they have a contribution to the value of the Louville form. The value of $\widetilde\theta_G$ on $[\phi,\psi]$ reads
$$\widetilde\theta_G([\phi,\psi])=\langle A,\;D_XY-D_YX+[X,Y]_\mg+D_BY-D_CX\rangle. $$
Finally, we can rewrite (\ref{C1}) and obtain the full expression for $\widetilde\omega_G(\phi,\psi)$
\begin{equation}\label{omega}
\widetilde\omega_G(\phi,\psi)=\langle B,Y\rangle-\langle C,X\rangle-\langle A,[X,Y]_\mg\rangle.
\end{equation}
Once we have a symplectic form on a the cotangent bundle to the group, we can find the map $\wtb_G$
$$\wtb_G:G\times\mathfrak g^*\times\mathfrak g\times\mathfrak g^*\to G\times\mathfrak g^*\times\mathfrak g^*\times\mathfrak g, \quad     v\longmapsto\widetilde\omega_G(\cdot,v) ,$$

\begin{equation}\label{beta}
\widetilde\beta_G(g,A,X,B)=(g,A,-B+\mathrm{ad}^*_X(A),X).
\end{equation}
Let us also write the inverse of (\ref{beta})
\begin{equation*}
\widetilde\beta_G^{-1}(g,A,B,X)=(g,A,X,-B+\mathrm{ad}^*_X(A)).
\end{equation*}
The map $\beta_G$ may be trivialized in an alternative way. According to formula  (\ref{gamma}) the bundles $\sT^*\sT^*G$ and $\sT^*\sT G$ are canonically isomorphic by $\gamma_G$. $\beta_G$ may be then defined as composition of $\alpha_G$ and $\gamma^{-1}_G$. We can proceed similarly in case of $\widetilde\beta_G$ by putting $\widetilde\beta_G=\widetilde\gamma^{-1}_{\sT G}\circ\widetilde\alpha_G$. We just have to find a trivialization of $\gamma_{\sT G}$. It turns out, that
$$\widetilde\gamma_{\sT G}: G\times\mathfrak g^*\times\mathfrak g^*\times\mathfrak g\to G\times\mathfrak g\times\mathfrak g^*\times\mathfrak g^*,\quad (g,A,B,X)\longmapsto (g,X,-B,A). $$
The composition $\widetilde\beta_G=\widetilde\gamma_G^{-1}\circ\widetilde\alpha_G$ gives precisely the formula (\ref{beta}).

\subsection{Tulczyjew triple on a Lie group}

The trivialized Tulczyjew triple is the following diagram
\begin{equation}\label{trojka}
\xymatrix@C-10pt{
G\times{\color{red}{\mathfrak g^*}}\times\mathfrak{g^*}\times \mathfrak{g} \ar[dr]^{pr_{12}} && G\times{\color{red}{\mathfrak g^*}}\times{\color{blue}{\mathfrak g}}\times \mathfrak{g^*}\ar[rr]^{\widetilde\alpha_G} \ar[ll]_{\widetilde\beta_G} \ar[dl]_{pr_{12}} \ar[dr]^{pr_{13}} && G\times{\color{blue}{\mathfrak g}}\times\mathfrak{g^*}\times \mathfrak{g^*} \ar[dl]_{pr_{12}}  \\
&G\times\color{red}{\mathfrak{g^*}} \ar[dr]^{pr_{1}}&& G\times\color{blue}{\mathfrak{g}} \ar[dl]_{pr_{1}}& \\
&&G&&}
\end{equation}
Using $\widetilde\alpha_G$ and $\widetilde\beta_G$ we can construct the dynamics from Lagrangian or Hamiltonian.

Let the configuration space of the system be a Lie group~$G$ and let $L:\sT G\rightarrow\mathbb{R}$ be the Lagrangian of the system. Composing $L$ with $\imath$ we can introduce a trivialized Lagrangian~$\widetilde L=L\circ\imath^{-1}$
$$\widetilde L:G\times\mathfrak g\to\mathbb R. $$
The trivialized dynamics of the system is a set
$$G\times\mathfrak g^*\times\mathfrak g\times\mathfrak{g^*}\supset\widetilde D:=\widetilde\alpha^{-1}_G\circ\mathrm{d}\widetilde L(G\times\mathfrak g),   $$
that may be written as
$$\mathsf d\widetilde L(g,X)=(g,X,\frac{\partial\widetilde L}{\partial g},\frac{\partial\widetilde L}{\partial X})\in G\times\mathfrak g\times\mathfrak g^*\times\mathfrak g^*,$$
\begin{equation}\label{d:1}
\widetilde{\mathcal D}=\left\{ (g,A,X,B): \quad B=\frac{\partial\widetilde L}{\partial X},  \quad A=\frac{\partial\widetilde L}{\partial g}+\mathrm{ad}^*_X(B)  \right\}.
\end{equation}
The trivialized dynamics is a Lagrangian submanifold in $G\times\mathfrak g^*\times\mathfrak g\times\mathfrak{g^*}$ with respect to $\mathrm{d}_{\sT}\widetilde\omega_G$. In case when $\widetilde L$ is regular, the dynamics is an image of the vector field on $G\times\mathfrak g^*$ otherwise it is an implicit differential equation on curves in $G\times\mathfrak g^*$.

Now, let $H:\sT^\ast G\to\mathbb R$ be a Hamiltonian of the system. Then, as for a Lagrangian, we can introduce the trivialized Hamiltonian $\widetilde H:G\times\mathfrak g^*\to\mathbb{R}$ composing $H$ with $\xi$. Using $\widetilde\beta_G$ mapping we define on $G\times\mathfrak g^*$ a Hamiltonian vector field of $\widetilde H$
$$\widetilde{X}_H:=\widetilde\beta^{-1}_G\circ\mathrm{d}\widetilde H,$$
that can be written as
$$\mathsf d\widetilde H(g,A)=(g,A,\frac{\partial\widetilde H}{\partial g},\frac{\partial\widetilde H}{\partial A}),$$
\begin{equation}\label{d:2}
\widetilde X_H(g,A)=(g,A,X,C), \qquad X=\frac{\partial\widetilde H}{\partial A}, \quad C=-\frac{\partial\widetilde H}{\partial g}+\mathrm{ad}^*_X(A).
\end{equation}

\section{Reduced Tulczyjew Triple}\label{sec:7}
As we have mentioned in previous section Tulczyjew triple can be reduced with respect to symmetries of the system. The result of the reduction is a triple based on a vector bundle which is not a tangent bundle to any manifold. What we usualy get is a triple based on an algebroid \cite{GGU}. In this section we shall assume that Lagrangian and Hamiltonian are invariant with respect to the group action on itself that is prolonged to $\sT G$ and $\sT^\ast G$. We start with reducing $\widetilde\alpha_G$, $\widetilde\beta_G$ and $\widetilde\kappa_G$ and then we shall use them to construct reduced Tulczyjew triple.

\subsection{Reduced involution of the iterated tangent bundle}

The group $G$ acts on the tangent bundle $\sT G$ by a tangent lift of left translation i.e. by $\sT l_g$. We can consider orbits of this action, identifying points on the same orbit. Using the isomorphism $\sT G/_G\simeq\mathfrak g$ we obtain
$$\tau: \sT G\to \mathfrak{g}.$$
Find the reduction of $\sT\sT G$ means applying the tangent functor to projection $\tau$
$$\sT\tau: \sT\sT G\to \sT\mathfrak{g}\simeq\mathfrak{g}\times \mathfrak{g}.$$
After trivialization both $\tau$ and $\sT\tau$ may be written as
\begin{eqnarray}
pr_2&:& G\times{\color{red}\mathfrak g}\to{\color{red}\mathfrak g}, \\
pr_{24}&:& G\times{\color{red}\mathfrak g}\times\mathfrak g\times{\color{blue}\mathfrak g}\to{\color{red}\mathfrak g}\times{\color{blue}\mathfrak g}, \label{r:1}
\end{eqnarray}
where $\tau$ corresponds to $pr_2$ and $\sT\tau$ corresponds to $pr_{24}$.
Once we have reduced the bundle $\sT\sT G$, we can try to reduce $\widetilde\kappa_G$. It is important to notice that the result of such a reduction may not be a map any more. Let us denote the reduced version of $\widetilde\kappa_G$ by $\kappa_{\mathfrak{g}}$.   Dividing by the group action in both source and target $\sT\sT G$ of $\kappa_G$  we see that the elements $(X_1,Y_1)\in\mathfrak g\times\mathfrak g$ and $(X_2,Y_2)\in\mathfrak g\times\mathfrak g$ are in relation $\widetilde\kappa_{\mathfrak g}$ if $Y_2=Y_1-[X_1,X_2]$. The image $\kappa_{\mathfrak g}(X_1,Y_1)$ contains more than one pair $(X_2,Y_2)$. The following diagram with is commutative in generalized sense of relations, not maps. Special arrow with triangle head distinguishes relation which is not a map.
$$\xymatrix@C+20pt{
G\times{\cob \mathfrak{g}}\times \mathfrak{g}\times{\cor\mathfrak{g}}\ar[r]^{\widetilde\kappa_G}\ar[d]^{pr_{24}} & G\times {\cob \mathfrak{g}}\times\mathfrak{g}\times{\cor\mathfrak{g}}\ar[d]^{pr_{24}} \\
{\cob \mathfrak g}\times {\cor\mathfrak{g}}\ar@{-|>}[r]^{\widetilde\kappa_{\mg}} & {\cob\mathfrak g}\times{\cor \mathfrak{g}} }
$$

\subsection{Reduced Tulczyjew isomorphism}

As we stated, the map $\widetilde\alpha_G$ acts between trivialized bundles $\sT\sT^\ast G$ and $\sT^\ast\sT G$. We can reduce the bundle \mbox{$\sT^\ast\sT G\simeq G\times\mathfrak g\times\mathfrak g^*\times\mathfrak g^*$} to $\mg\times\mg^*$ by means of symplectic reduction with respect to certain coisotropic submanifold $K\subset\sT^\ast\sT G$. In trivialisation $K$ becomes $\widetilde K$
$$\widetilde K:=\{(g,X,0,A)\in G\times\mathfrak g\times\mathfrak g^*\times\mathfrak g^*    \} $$
and consists of differentials of functions on $G\times\mathfrak g$ invariant with respect to the action of $G$ on first component. Symplectic reduction coincides with dropping the first element in $(g,X,0,A)$ and forgetting about $0$ on the third place
\begin{equation}\label{r:6}
\xymatrix{
G\times{\color{red}\mathfrak{g}}\times \mathfrak{g^*}\times{\color{blue}\mathfrak{g^*}}\supset\widetilde K \ar[r]^-{pr_{24}}    &
{\color{red}\mathfrak{g}}\times {\color{blue}\mathfrak{g^*}}}
\end{equation}
Group $G$ acts on $\sT^\ast G$ by the cotangent lift of the left action on $G$. If we divide the cotangent bundle by this action we get (in trivialization)
$$pr_{2}: \sT^*G\simeq G\times\mathfrak g^*\to\mathfrak g^*.$$
Applying the tangent functor to this map we obtain
\begin{equation}\label{r:61} pr_{24}: G\times{\color{red}\mathfrak g^*}\times\mathfrak g\times{\color{blue}\mathfrak g^*}\to{\color{red}\mathfrak g^*}\times{\color{blue}\mathfrak g^*}.\end{equation}
Reduction of $\widetilde\alpha_G$ should be in some sense composed of $\widetilde\alpha_G$ and both above reductions (\ref{r:6}, \ref{r:61}).
The fiber of $pr_{24}$ in (\ref{r:61}) over $(A,B)$ contains all elements of $G\times\mathfrak{g^*}\times \mathfrak{g}\times\mathfrak{g^*}$ of a form $(g,A,X,B)$. Acting by $\widetilde\alpha_G$ we get
$$\widetilde\alpha_G(g,A,X,B)=(g,X,B-\text{ad}^*_X(A),A).$$
Element $(g,X,B-\text{ad}^*_X(A),A)$ to belongs to $\widetilde K$ if $B=\mathrm{ad}^*_X(A)$. Then we can apply $pr_{24}$ from (\ref{r:6}).

Finally, a pair $(A,B)$ is in relation with a pair $(X,A)$ if $B=\mathrm{ad}^*_X(A)$. In particular it means, that there is no proper $(X, A)$ for each element $(A,B)$. However, one can easily notice the opposite: for each pair $(X,A)$ there exist an element $(A,B=\mathrm{ad}^*_X(A))$ that is in relation with $(X,A)$. Thus, the relation $\alpha_{\mg}$ is a map `in the opposite direction' comparing to the original map $\widetilde\alpha_{G}$:
\begin{equation}\label{r:5}
\alpha_{\mg}: \mg\times\mg^\ast\to\mg^\ast\times\mg^\ast (X,A)\longrightarrow (A, \mathrm{ad}^*_X(A)).
\end{equation}
The appropriate diagram is
\begin{equation}\label{r:7}
\xymatrix{
G\times{\color{red}\mathfrak{g^*}}\times \mathfrak{g}\times {\color{blue}\mathfrak{g^*}} \ar[r]^{\widetilde\alpha_G} \ar[dd]^{pr_{24}} &
  G\times{\color{red}\mathfrak{g}}\times \mathfrak{g^*}\times{\color{blue}\mathfrak{g^*}} \\
 &\widetilde K \ar@{ (->}[u] \ar[d]^{pr_{24}} &   \\
{\color{red}\mathfrak{g^*}}\times {\color{blue}\mathfrak{g^*}}   & {\color{red}\mathfrak{g}}\times {\color{blue}\mathfrak{g^*}}\ar[l]_{\alpha_{\mathfrak{g}}}}
\end{equation}

\subsection{Reduced mapping related to a symplectic form}

The $\widetilde\beta_G$ mapping acts between bundles $G\times\mathfrak g^*\times\mathfrak g\times\mathfrak g^*$ and $G\times\mathfrak g^*\times\mathfrak g^*\times\mathfrak g$, i.e. trivialized versions of $\sT\sT^\ast G$ and $\sT^\ast\sT^\ast G$ respectively. The first of these bundles has already been reduced in the previous subsection. The bundle $\sT^\ast\sT^\ast G\simeq G\times\mathfrak g^*\times\mathfrak g^*\times\mathfrak g$ reduces to $\mg^*\times\mg$ be means of a symplectic reduction with respect to certain coisotropic submanifold $C\subset\sT^\ast\sT^\ast G$. In trivialisation
$$\widetilde C:=\{(g,A,0,X)\in G\times\mathfrak g^*\times\mathfrak g^*\times\mathfrak g \} $$
and consists of differentials of functions on $G\times\mathfrak{g}^\ast$ invariant with respect to the action of $G$ on the first component. Symplectic reduction coincides with dropping the first element in $(g,A,0,X)$ and forgetting zero on the third place.
\begin{equation}\label{r:8}
\xymatrix{
G\times{\color{red}\mathfrak{g}^\ast}\times \mathfrak{g^*}\times{\color{blue}\mathfrak{g}}\supset
\widetilde C  \ar[r]^-{pr_{24}}   &
{\color{red}\mathfrak{g}^\ast}\times {\color{blue}\mathfrak{g}}}
\end{equation}
As in case of $\alpha_{\mathfrak{g}}$, applying $\wtb_G$ to $(g,A,X,B)$ we get an element of $\widetilde C$ if $-B+\text{ad}^*_X(A)=0$. The appropriate diagram
\begin{equation}\label{r:10}
\xymatrix{
G\times{\color{red}\mathfrak{g^*}}\times \mathfrak{g}\times {\color{blue}\mathfrak{g^*}} \ar[r]^{\widetilde\beta_G} \ar[dd]^{pr_{24}} &
  G\times{\color{red}\mathfrak{g}^\ast}\times \mathfrak{g^*}\times{\color{blue}\mathfrak{g}} \\
 &\widetilde C \ar@{ (->}[u] \ar[d]^{pr_{24}} &   \\
{\color{red}\mathfrak{g^*}}\times {\color{blue}\mathfrak{g^\ast}}   & {\color{red}\mathfrak{g}^\ast}\times {\color{blue}\mathfrak{g}}
\ar[l]_{\beta_{\mathfrak{g}}} }
\end{equation}
The reduced relation $\beta_{\mathfrak g}$ is again a map `in the opposite direction' comparing to the original $\widetilde\beta_G$.
\begin{equation}\label{r:9}
\beta_{\mg}: \mg^\ast\times\mg\to\mg^\ast\times\mg^\ast,\quad (A, X)\longrightarrow (A,\mathrm{ad}^*_X(A)).
\end{equation}
Note that $\beta_\mg$ is associated to the canonical geometric structure on $\mathfrak{g}^\ast$ namely linear Poisson bivector. It is in the full agreement with the fact that reducing phase space with respect to symmetries usually leads from symplectic to Poisson structure.

\subsection{Reduced Tulczyjew triple}

The reduced Tulczyjew triple is the following diagram
\begin{equation}
\xymatrix{
&{\color{red}\mathfrak{g^*}}\times{\color{blue}\mathfrak{g}} \ar[dr]_{pr_1} \ar[dl]_{pr_2} \ar[r]^{\wtb_\mg}& {\color{red}\mathfrak{g^*}}\times \mathfrak{g^*}  \ar[d]_{pr_1} & {\color{blue}\mathfrak{g}}\times {\color{red}\mathfrak{g^*}} \ar[dl]^{pr_2} \ar[dr]^{pr_1}  \ar[l]_{\wta_\mg}& \\
{\color{blue}\mathfrak{g}} & & {\color{red}\mathfrak{g^*}} & &{\color{blue}\mathfrak{g}} \\
}
\end{equation}
It may be used to describe systems with Lagrangian and Hamiltonian invariant under the group action. The right-hand side represents the reduced Lagrangian formalism while the left-hand side represents the reduced Hamiltonian formalism.

Let $L$ be a Lagrangian of the system defined on $\sT G$ and invariant under the group action
\begin{equation}\label{r:614}
L(v)=L(\sT l_g(v))  \quad g\in G, v\in \sT G.
\end{equation}
Since $L$ is invariant, we can introduce the reduced Lagrangian
\begin{center}
$L(v_g)=L(\sT l^{-1}_g(v_g))=:l(X)  \quad X\in\mathfrak g, \quad l: \mathfrak{g}\to \mathbb{R}. $
\end{center}
The reduced dynamics is the set
$$d=\alpha_{\mathfrak g}\circ\mathrm{d}l(\mathfrak g)\subset\mathfrak g^*\times\mathfrak g^*,$$
i.e.
\begin{equation}\label{r:11}
d=\{ (A,B)\in\mathfrak{g^*}\times\mathfrak{g^*}:\quad \exists X\in \mg \quad A=\frac{\partial l}{\partial X},\quad B=\mathrm{ad}_X^*(\frac{\partial l}{\partial X})\,\}.
\end{equation}

Now let us assume that the Hamiltonian of the system does exists and is invariant under the group action i.e. $H(b)=H(\sT l_g^*(b))$ for each $b\in\sT^*G$.
Then we can introduce the reduced Hamiltonian $h$
$$H(b_g)=H(\sT l_g^*(b_g))=:h(B) \quad B\in\mathfrak g^*,\quad h:\mathfrak g^*\to \mathbb R. $$
The Hamiltonian vector field of $h$ is given by $X_h:=\beta_{\mg}(\mathrm dh)$
\begin{equation}\label{r:12}
X_h: \mg^*\to\mg^*\times\mg^*,\quad B\longmapsto (B,\mathrm{ad}^\ast_{\frac{\partial h}{\partial B}}(B)).
\end{equation}
One can also show, that the reduced Legendre map combining the Lagrangian and the Hamiltonian side has a form
\begin{equation}\label{r:13}
\lambda_{\mg}:\mg\to\mg^*,\quad X\longmapsto\frac{\partial l}{\partial X}.
\end{equation}
If the Legendre map is an isomorphism, then as in the non-reduced case, the dynamics $d$ is an image of the hamiltonian vector field $X_h$ of a proper reduced Hamiltonian $h$
$$h(B)=\langle B, \lambda_\mg^{-1}(B)\rangle-l(\lambda_\mg^{-1}(B)).$$
All elements of reduced mechanics may be put into the diagram of reduced Tulczyjew triple
$$
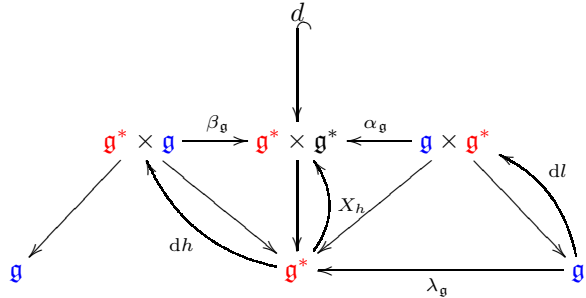
\xymatrix@R+10pt{
&& d \ar@{^{(}->}[d] && \\
&{\color{red}\mathfrak{g^*}}\times{\color{blue}\mathfrak{g}} \ar[dr] \ar[dl] \ar[r]^{\beta_\mg}& {\color{red}\mathfrak{g^*}}\times \mathfrak{g^*}  \ar[d]  & {\color{blue}\mathfrak{g}}\times {\color{red}\mathfrak{g^*}} \ar[dl] \ar[dr]  \ar[l]_{\alpha_\mg}& \\
{\color{blue}\mathfrak{g}}  &&{\color{red}\mathfrak{g^*}} \ar@/^1pc/[ul]^{\mathrm{d}h} \ar@/_1pc/[u]_{X_h} && {\color{blue}\mathfrak{g}} \ar@/_1pc/[ul]_{\mathrm{d}l} \ar[ll]^{\lambda_\mg}
}
$$

\section{Example}\label{sec:8}

We will present now an application of the reduced Tulczyjew formalism for a particular physical system which is a rigid body fixed at one point and free to rotate about it.

Let us fix the initial position of the rigid body in $\mathbb{R}^3$. Then any other position can be identified with an element $R$ in $G=SO(3)$ and any movement can be described by a curve $t\mapsto R(t)$ in $G$. Trajectory of a point $q\in \mathbb{R}^3$ of the body is given by a curve $t\mapsto R(t)q$ where we consider the natural action of $SO(3)$ on $\mathbb{R}^3$. Let
$(\cdot|\cdot)$ denote the canonical Euclidean scalar product on $\mathbb{R}^3$. Lagrangian of a rigid body reads
\begin{equation}\label{lag}
L(\dot{R})=\frac{1}{2}\int_{V} \rho (q)(\dot{R}q|\dot{R}q) d^3q,
\end{equation}
where $\rho$ is a mass density function of the rigid body, $\dot{R}q$ denotes vector tangent to $t\mapsto R(t) q\in \mathbb{R}^3$ at $t=0$ and integration is over the whole volume of the rigid body with respect to the canonical density $d^3q$ on $\mathbb{R}^3$.

The Lagrangian (\ref{lag}) is a function on $\sT SO(3)$. One can easily see that it is invariant under the transformation (\ref{r:614}). It means that we can introduce the reduced Lagrangian
\begin{equation}
l:\mathfrak{so}(3)\to\mathbb R, \qquad l(X)=\cfrac{1}{2}\int \rho (q)(Xq|Xq) d^3q,
\end{equation}
where $Xq$ denotes vector tangent to the curve $t\mapsto \exp(tX)q\in\mathbb{R}^3$. Let us note that the integral in above formula depends in a bilinear and symmetric way on $X$, therefore it defines a bilinear symmetric form on $\mathfrak{so}(3)$ called {\it moment of inertia}. Let us denote this form by $I$ i.e $I$ is given by
\begin{equation}\label{tnsr}
I(X,Y)=\int_V \rho (q)(Xq|Yq)\,d^3q
\end{equation}
Due to nondegeneracy it defines an isomorphism
\begin{equation}\label{tnsr:1}
\tilde I:\mathfrak{so}(3)\to\mathfrak{so}^*(3),\quad X\longmapsto I(X,\cdot).
\end{equation}
We can now rewrite $l$ using (\ref{tnsr})
$$l(X)=\cfrac{1}{2}I(X,X).$$
Once we have $l$ we can easily find the reduced dynamics. Using the formula (\ref{r:11}) we obtain
\begin{equation}\label{dyn}
d=\{ (A,\dot A)\in\mathfrak{so}(3)^*\times\mathfrak{so}(3)^*:\quad\exists X\in\mathfrak{so}(3)\;\; A=\tilde I(X),\;\;\dot A=\mathrm{ad}_{X}^*\tilde I(X) \}.
\end{equation}
Since $\tilde I$ is an isomorphism we can rewrite (\ref{dyn}) as
\begin{equation}\label{dyn:1}
d=\{ (A,\dot A)\in\mathfrak{so}(3)^*\times\mathfrak{so}(3)^*: \dot A=\mathrm{ad}_{\tilde I^{-1}(A)}^*A \}.
\end{equation}
It is easy to see that $d$ is an image of the Hamiltonian vector field on $\mathfrak{so}^*(3)$ with respect to the canonical Poisson structure for Hamiltonian function
\begin{equation}\label{dyn:2}
h:\mathfrak{so}^*(3)\to\mathbb R, \qquad h(A)=\frac{1}{2}\langle A, \tilde I^{-1}(A)\rangle.
\end{equation}
Let us notice that $\tilde I$ is the Legendre map for Lagrangian $l$, $A=\tilde I(X)$ is the canonical momentum and $l$ is hiperregular.

Since $\tilde I$ is an isomorphism, we can transport Hamiltonian vector field from $\mathfrak{so}^*(3)$ to $\mathfrak{so}(3)$ and get
\begin{equation}\label{PE}
\tilde I(\dot X)=\mathrm{ad}_{X}^*(\tilde I(X)).
\end{equation}
Now we identify $(\mathfrak{so}(3), [\cdot, \cdot])$  with $(\mathbb{R}^3,\times)$ denoting by $\vec X$ the element of $\mathbb{R}^3$ corresponding to $X$. Moreover, the dual space to $\mathbb R^3$ may be as well identified with $\mathbb R^3$ by means scalar product. The resulting $\tilde I$ will be denoted by $\bar I$. Equation (\ref{PE})  assumes the traditional form of {\textit{Euler equation}} \cite{HMR}
$${\bar I}(\dot{\vec{X}})-{\bar I}(\vec{X})\times \vec{X}=0.$$
Therefore, the result obtained with the reduced Tulczyjew triple is consistent with one derived in `traditional' way.

\end{document}